\documentstyle[preprint,aps,floats,epsfig]{revtex}
\tighten
\begin{document}
\draft
\preprint{MIT-CTP-3361}
\topmargin=0.1cm

\title{Tipping the balances of a small world}
\author{Lu\'{\i}s M. A. Bettencourt}
\address{Center for Theoretical Physics,
Massachusetts Institute of Technology,
Cambridge MA 02139 \\ } 
\address{Los Alamos National Laboratory, CCS-3 MS B256, 
Los Alamos NM 87545}

\date{\today}
\maketitle

\begin{abstract}
Recent progress in the large scale mapping of social networks is 
opening new quantitative windows into the structure of human societies. 
These networks are largely the result of how we access and utilize 
information. Here I show that a universal decision mechanism, where 
we base our choices on the actions of others, can explain much of their 
structure. Such collective social arrangements emerge from successful 
strategies to handle information flow at the individual level. 
They include the formation of closely-knit communities and the 
emergence of well-connected individuals. The latter can command the 
following of others while only exercising ordinary judgment. 
\end{abstract}

\pacs{PACS Numbers: 89.75.Fb, 89.75.-k, 89.65.Ef, 89.75.Hc}

In recent years there has been growing interest in the quantitative 
structure of human societies. It has emerged that we are part of heterogeneous 
networks or graphs \cite{1,2,3,4}, sets of links that connect each one of us to all our 
acquaintances. Not all people are alike: some live almost isolated, most belong 
to distinguishable communities \cite{1,5} and a small fraction of the population 
is made up of exceptionally well connected individuals \cite{6}. Social networks 
have the remarkable property that one can reach anyone else through a very 
small number of connections - the famous six degrees of separation \cite{7,8}. 

These findings beg important questions: Why are social networks invariably 
clustered in communities?  Why are there individuals with such different 
connectivity? Answering these puzzles requires tying the morphology of social 
networks to their function \cite{2,9,10}.  Similar problems occur in the study of 
other complex networks, for example, dealing with gene and protein-protein 
interactions \cite{11,12,13}, metabolism \cite{14,15}, ecosystems \cite{16,17} (foodwebs) 
and neural activity. Thus understanding the simultaneous robustness and 
adaptability of these complex systems in the light of their function is a 
general problem at the forefront of the current scientific agenda across 
many disciplines \cite{9}. 

The difficulty of this approach consists in defining the function of each of 
these complex networks in a way that captures their essence and simultaneously 
permits quantitative progress. Clearly many details of social behavior, in 
particular, appear too rich and our understanding of them remains too 
qualitative to fall in this class. There are however important well documented 
exceptions. 

A familiar situation is having to choose between seemingly equivalent options, 
at least given the amount of information and time at our disposal \cite{18}. 
In practice many of our decisions fall in this class. This leads to a 
degeneracy of choice, typical also of situations when relevant information 
is difficult to discriminate from too much noise, or when it cannot be trusted. 
In these situations we often rely on the observation of the actions of others 
we know as the basis for our decisions \cite{19,20,21,22,23}. This strategy has two important 
advantages: we can be sure not to do worse than most of the people we know and, 
in addition, we may actually join a winning trend early and profit from it.

Recently this type of discriminating imitation has become the focus of an 
extensive empirical literature in economy \cite{19} and the social sciences. 
Bikhchandani, Hirshleifer and Welch \cite{20,21} collected a vast amount of 
empirical evidence that establishes the universal importance of the 
choices of others in influencing our own and were able to model this 
phenomenon in simple terms. They dubbed the formation of the trend 
or fad that often results an information cascade; a process whereby 
sequential individual choices propagate a piece of information through 
the entire population \cite{22}.  This phenomenon is also often liked 
(qualitatively) to the spread of an epidemic \cite{10,22}. Interestingly, 
information cascades lead to the spontaneous formation of large consensus 
where there are a priori no individual preferences. 

Here I use an implementation of these ideas \cite{23} consisting of a population 
of $N$ agents, facing a choice among $L$ labels. At each time step individuals 
compare the relative growth rate of their label to that of one of their 
immediate acquaintances', chosen at random. If the latter's growth rate (the trend's 
relative momentum) is greater the agent switches to his neighbor's trend; otherwise 
he keeps his. The model has one additional ingredient: if a trend slows down 
individuals may decide to take a risk in something new (an empty label). 
This effect is modeled by $p_{\rm crit}$, the relative growth rate below which 
non-conformism sets in. Here $p_{\rm crit}=10^{-5}$, which results in population 
wide trends or cascades \cite{23}. References to other related dynamical 
implementations \cite{22} (including  models of herding) and additional discussion 
are given elsewhere \cite{23}. 

Typical dynamics \cite{23} are characterized by cycles alternating population 
disorder, when many different trends coexist, and order, when most of the 
population falls into the same label. Both collective states of order and 
disorder are dynamically unstable making the evolution very sensitive to chance 
events \cite{20,21,23}. As a result it becomes very difficult in practice for an 
external observer to profit from the reckoning that agents are following trends, 
especially when the number of choices becomes large.

To explore the effects of the underlying network morphology on the dynamics I generate 
(binary) artificial social networks as small world graphs \cite{1,24}. These are random 
graphs with clustering: $N$ individuals are represented as nodes, each with an 
average number of connections $z$. Clustering is produced by dividing the population 
into communities, each characterized by an average higher degree of internal connections 
per node $z_{\rm in}$  than external $z_{\rm out}$ connections 
($z=z_{\rm in}+z_{\rm out}$).   

\begin{figure}
\begin{center}
\epsfig{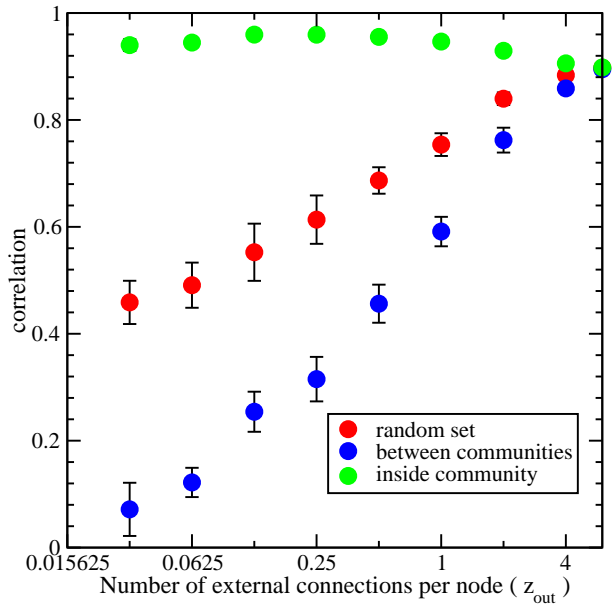}
\vskip 0.5 cm
\caption{The correlation between two halves of the same community (green), 
halves of distinct communities (blue) and a set of individuals with random 
connections (red) for $N=256$, $L=1000$, $z=8$, divided in 4 communities (see text). 
The correlation inside a community is always close to 1. The correlation and 
synchronization of choices between distinct communities is low for small $z_{\rm out}$, 
becoming higher as the number of connections between them increases.  
Individuals with random connections display intermediate correlation. 
For high $z_{\rm out}$ the original communities merge together. Error bars denote 
standard deviations over a set of 20 network realizations and many cascade cycles. }
\label{fig1}
\end{center}
\end{figure}

In addition to measure correlations between parts of the population it is 
useful to define a label state vector 
\begin{eqnarray}
{\vec v}_c = \left( N^c_1, N^c_2,\ldots,N^c_L \right)/ A_c, \qquad A_c = \sqrt{\sum_{i=1}^L 
\left( N^c_i \right)^2 },
\end {eqnarray}                                           
where $N^c_i$ denotes the number of individuals in label $i$ belonging to group $c$.  
The natural inner product
\begin{eqnarray}
\langle {\vec v}_a \vert {\vec v}_b \rangle = \left( A_a A_b \right)^{-1} \sum_{i=1}^L 
\left( N^a_i N^b_i \right)
\end{eqnarray}                                 
is a (positive definite) measure of the correlation between different groups, 
see Fig.\ref{fig1}.

Fig.~\ref{fig1} shows the correlation between several subsets of the population, within 
a community, between two distinct communities and for a control set of individuals with 
random connections. It is now clear why it is a good defensive strategy to belong to 
a tightly knit collective: communities are {\it islands of information coherence}. 
Thanks to the large redundancy of personal connections inside the community the 
coherence of local information is preserved over time and personal deviations inside 
the group are small compared to those to the outside. This remains true even if a 
few individuals or connections are lost. 

\begin{figure}
\begin{center}
\epsfig{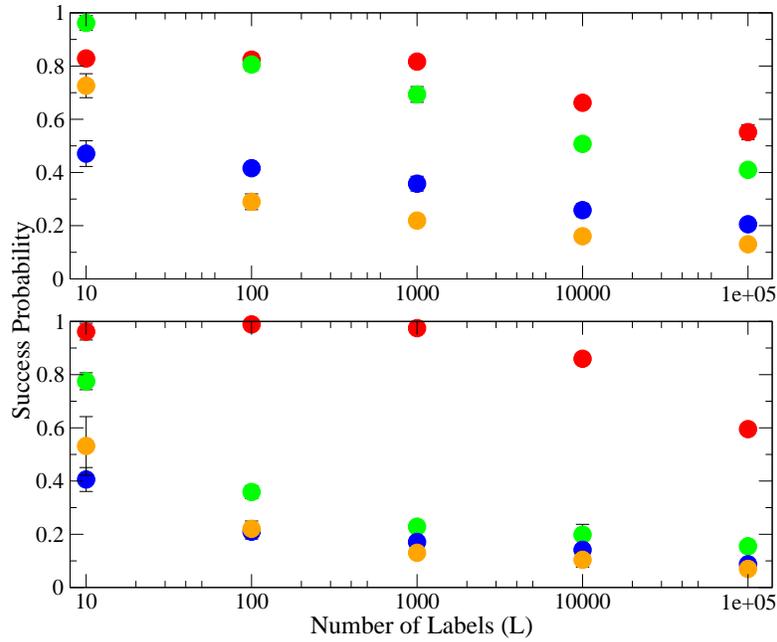}
\vskip 0.5 cm
\caption{The success rate of several criteria for predicting the next winning 
trend at the time when the dominant movement decays. The next winning trend is not 
easily determined as the largest secondary trend (blue), the fastest growing (orange) or 
even the trend with the largest product of size and momentum (green). The best predictor 
is the choice of the hub (red), particularly as the number of choices $L$ becomes large. 
The upper panel refers to lower hub visibility (his input is considered on average 
by each individual with probability $p_{\rm hub}=1/8$, each time), the lower panel 
to higher visibility ($p_{\rm hub}=1/2$). Error bars are as in Fig.~\ref{fig1}.}
\label{fig2}
\end{center}
\end{figure}

Comforting as it may be to keep up with our neighbors it may actually be better to 
be a step ahead. As we discussed above this is a tall order, even if one is fully 
informed of the state of the whole population. Figure~\ref{fig2} shows the success rate 
of several criteria attempting to predict the emerging new trend at the particularly 
important time when a former dominant movement collapses, i.e. when it becomes as large as 
the largest secondary trend. All criteria based on the full knowledge of the state of 
population at this particular time (the largest secondary trend, the fastest growing 
one or the trend with the largest product of momentum and size) are far from good 
and become very poor for large number of competing choices L.

\begin{figure}
\begin{center}
\epsfig{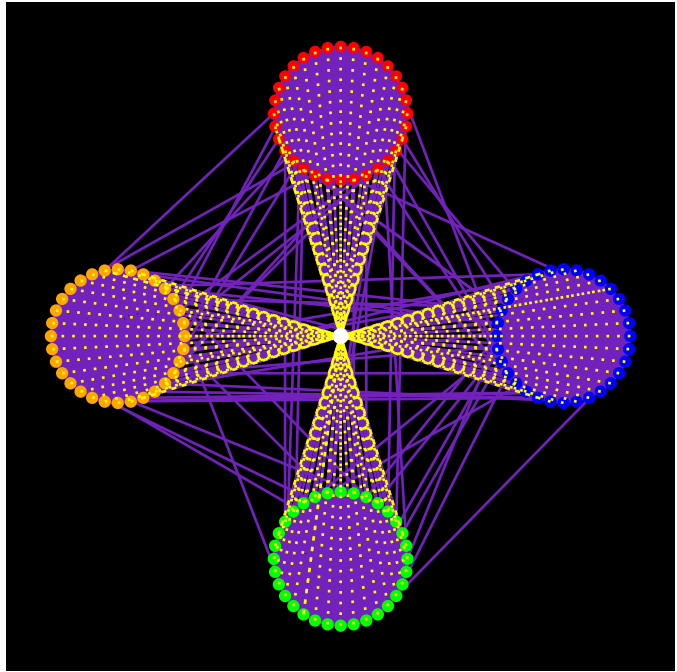}
\vskip 1 cm
\caption{An example of a (binary) social network with $N=128$, $z=4$, divided 
into 4 communities (red, blue, green, orange) and with a hub (central node). 
Here the state of the hub is seen by all individuals with probability $p_{\rm hub}=0.25$ 
each time, but has input from $z=4$ individuals. As such the actions of the hub are 
very visible but not better informed.}  
\label{fig3}
\end{center}
\end{figure}

\begin{figure}
\begin{center}
\epsfig{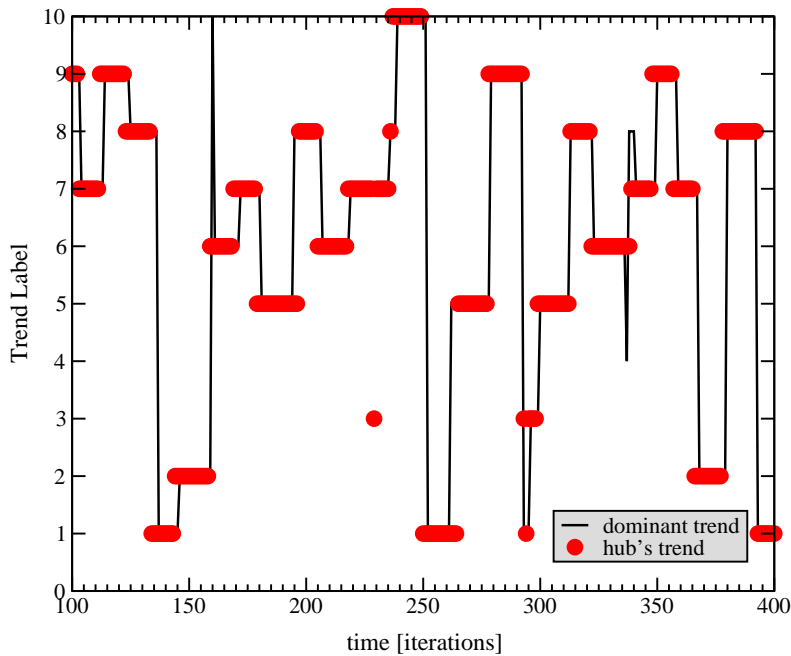}
\vskip 0.5 cm
\caption{Fig. 4 Evolution of the dominant trend (solid line) and the hub's trend 
(red circles) for $L=10$. The hub invariably picks the next dominant trend correctly 
and early, qualities that reinforce his social role as a bellwether.}   
\label{fig4}
\end{center}
\end{figure}

Interestingly there is a simple alternative solution - it relies on connections, not 
reasoning or information. I examine this scenario by introducing a new well-connected 
individual into the population, a network hub, as in Fig.~\ref{fig3}. The hub bases 
his decisions, like any other agent, on the state of an average number of other 
individuals $z$, but his choices can be seen by everybody else. What is particular 
about the evolution is illustrated in Figs.~\ref{fig2} and \ref{fig4}: the hub is 
exceptionally good at picking the next winning trend early, before it becomes 
dominant - the perfect winning strategy.

However the hub is, by construction, neither better informed nor animated by superior 
decision making.  This apparent paradox is easily dispelled: the hub's actions are 
very visible to others. Any reasonable decision on his part (the adoption of any 
growing label) has a large probability of being immediately followed 
by many ($\sim N p_{\rm hub}$, at the next iteration) 
and thus to {\it make} the winning trend. This property is independent of the underlying 
community structure and is enhanced for larger populations (larger $N$), as long 
as $p_{\rm hub}(N)$ is such that $d(N p_{\rm hub})/dN>0$. Thus, it is popularity, 
not knowledge or reasoning, that leads to the most successful strategy in an 
environment characterized by strong choice degeneracy.

Given some memory the hub's successes reinforce his position and (apparent) foresight. 
Each correct 'prediction' encourages others to heed his choices and follow at 
the next opportunity. This reinforces the hub's popularity, allowing him to 
pick the next winning trend with greater certainty and so on: the process 
is self-reinforcing. It also naturally leads to a specific form of preferential 
attachment \cite{25}, where the most connected node - the best trend predictor 
- is preferred. Thus, under choice degeneracy, one should expect the appearance of 
well-connected, very visible individuals as a social network evolves.

Nodes with an exceptionally large connectivity are a common property of 
other complex networks, including scale-free graphs \cite{26} describing e.g. 
WWW, power-grids and the large scale features of protein-protein interactions 
{\it in vivo}. It has been pointed \cite{27,28} out that such networks' utmost fragility 
is due to the loss of these key nodes.  Trend dynamics shows how this fragility 
may only be apparent in social networks. The hub is a common node, only its degree 
of outgoing connections is exceptionally large. I argued above that there is a 
fundamental instability for a common individual to be promoted to this position. 
Because social connections are rearranged on much faster time scales than nodes \cite{29}, 
upon loss of a hub a new one can quickly develop from another node and the structural 
integrity of the network will be preserved after a short transient. Moreover the 
addition of a second well-connected node dealing with the same information reduces 
the predictability of emerging trends, unless the two hubs work in tandem 
(as would happen under specific assortative mixing \cite{29}) and so forth. 
It is however perfectly natural for separate hubs to coexist if they relate 
to different social dimensions \cite{30}, i.e. if they deal with different types of 
information. In this way the large scale structure of human societies, when 
averaged over time and social dimensions may be characterized by many hubs with 
varying reaches and interdependencies. These properties may lead to the emergence 
of interesting scale-invariances in large social networks associated with decision 
making and information flow. 

Observing the actions of others is a universal simple mechanism that allows us 
to handle imperfect information in our complex social environment to make difficult 
decisions. We can protect ourselves from the tyranny of fashions by associating 
into tightly knit communities or we may try to set trends by influencing the choices 
of others through our social connections. Here I showed that these successful individual 
strategies lead to stable social arrangements, which coincide with some of the 
most notable observed structures of social networks. Trend dynamics breaks the 
degeneracy of our individual choices and leads to the spontaneous formation of 
collective movements. Whenever concerted social action is more productive than 
the sum of individual efforts social hubs may become the social mechanism 
that facilitates the creation of consensus most promptly and predictably.

\section*{Acknowledgments}

Acknowledgments: I would like to thank N. D. Antunes, J.-P. Bouchaud, D. Kaiser, 
L. M. Rocha and B. Svetitsky for useful discussions. This work is supported in 
part by the Department of Energy under cooperative research agreement 
$\#$DF-FC02-94ER40818.


\begin{thebibliography}{99}

\bibitem{1} D. J. Watts and S. H. Strogatz, Collective dynamics of small world networks, 
Nature {\bf 393}, 440-442 (1998).  

\bibitem{2} D. J. Watts, Small worlds (Princeton University Press, Princeton University 
Press, Princeton, 1999).

\bibitem{3} L. A. N. Amaral,  A. Scala, M. Barthelemy, H. E. Stanley, Classes of 
small-world networks, Proc. Natl. Acad. Sci. U.S.A. {\bf 97}, 11149-11152 (2000).  

\bibitem{4} E. M. Jin, M. Girvan, M. E. J. Newman, Structure of growing social networks, 
Phys. Rev. E {\bf 64}, 046132 (2001).  

\bibitem{5} M. Girvan and M. E. J. Newman, Community Structure in social and biological networks, 
Proc. Natl. Acad. Sci. USA {\bf 99}, 8271-8276 (2002). 

\bibitem{6} M. Gladwell, The tipping point: How little things can make a big difference 
(Little, Brown and Company, Boston, 2000). 

\bibitem{7} S. Milgram,  The small world problem, Psychol. Today {\bf 2}, 60-67 (1967).

\bibitem{8} I. Pool and M. Kochen,  Contact and influence, Social Networks {\bf 1}, 1-48 (1978). 

\bibitem{9} L. H. Hartwell, J. J. Hopfield, S. Leibler, A. W. Murray, From molecular to 
modular cell biology, Nature {\bf 402}, 47-52 (1999).

\bibitem{10} S. H. Strogatz, Exploring complex networks, Nature {\bf 410}, 268-276 (2001).

\bibitem{11} H. Jeong, S. P. Mason, Z. N. Oltvai, A.-L. Barab\'asi, Lethality and centrality in 
protein networks, Nature  {\bf 411}, 41-42 (2001).

\bibitem{12} S. Maslov and K. Sneppen, Specificity and Stability in Topology of Protein 
Networks, Science {\bf 296}, 910-913 (2002). 

\bibitem{13} A.-C. Gavin,  {\it et al.} 
Functional organization of the yeast proteome by systematic 
analysis of protein complexes, Nature {\bf 415}, 141-147 (2002).

\bibitem{14} H. Jeong, B. Tombor, R. Albert, Z. N. Oltvai, A.-L. Barab\'asi, The large-scale 
organization of metabolic networks, Nature  {\bf 407}, 651-654 (2000).

\bibitem{15} D. A. Fell and A. Wagner, The small world of metabolism, Nature Biotechnology 
{\bf 18}, 1121-1122 (2000).    

\bibitem{16} S. L. Pimm, J. H. Lawton, J. E. Cohen, Food web patterns and their consequences, 
Nature {\bf 350}, 669-674 (1991).

\bibitem{17} R. J. Williams and N. D. Martinez,  Simple rules yield complex food webs, 
Nature {\bf 404}, 180-183 (2000).  

\bibitem{18} This degeneracy of choice was first formalized by J. Bernouilli, Ars 
Conjectandi (Thurnisiorum, Basil, 1713) as 'the principle of insufficient reason'; 
later it was also known as 'the principle of indifference', J. M. Keynes, 
A Treatise on probability (Macmillan, London, 1921). It is also a well-known criterion
in the context of the theory of decision making under uncertainty, prescribing 
that a decision should be reached under the assumption of a uniform probability 
distribution over 'states of nature'.  

\bibitem{19} The importance of the decisions of others in setting our own choices was 
first articulated in print by J. M. Keynes in The general theory of employment, 
interest and money  (Macmillan, London, 1936), pages 155-156.

\bibitem{20} S. Bikhchandani, D. Hirshleifer,  I. Welch, A theory of fads, fashion, 
custom, and cultural change as information cascades, J. Pol. Economy {\bf 100}, 992-1026 (1992).

\bibitem{21} S. Bikhchandani, D. Hirshleifer,  I. Welch, Learning from the behavior 
of others: conformity, fads and information cascades, J. Econ. Perspectives {\bf 12}, 151-170 (1998).

\bibitem{22} See also D. J. Watts, A simple model of global cascades on random networks, 
Proc. Natl. Acad. Sci. U.S.A. {\bf 99}, 5766-5771 (2002), which explores similar dynamics 
over random networks. 

\bibitem{23} L. M. A. Bettencourt, From boom to bust and back again: The complex dynamics 
of trends and fashions, (available at http://arXiv.org/abs/cond-mat/0212267).

\bibitem{24} M. E. J. Newman,  Models of the small world, a review, 
J. Stat. Phys. {\bf 101}, 819-841 (2000).

\bibitem{25} A.-L. Barab\'asi and R. Albert, Emergence of scaling in random networks, 
Science {\bf 286} 509-512 (1999).

\bibitem{26} R. Albert and A.-L. Barab\'asi, Statistical mechanics of complex networks, 
Rev. Mod. Phys. {\bf 74}, 47-97 (2002).

\bibitem{27} R. Albert, H. Jeong, A.-L. Barab\'asi, Error and attack tolerance of complex 
networks, Nature {\bf 406}, 378-381 (1999).

\bibitem{28} D. S. Callaway, M. E. J. Newman, S. H. Strogatz, D. J. Watts,  Network robustness and 
fragility: Percolation on random graphs, Phys. Rev. Lett. {\bf 85}, 5468-5471 (2000). 

\bibitem{29} M. E. J. Newman, Mixing patterns in networks: Empirical results and models, 
(available at http://arXiv.org/abs/cond-mat/0209450); M. E. J. Newman and 
M. Girvan, Mixing patterns and community structure in networks, 
(available at http://arXiv.org/abs/cond-mat/0210146). 

\bibitem{30} D. J. Watts, P. S. Dodds, M. E. J. Newman, Identity and search in social 
networks, Science {\bf 296}, 1302-1305 (2002).  




\end{thebibliography}
\end{document}